\documentclass[twocolumn,showpacs,preprintnumbers,amsmath,amssymb]{revtex4}
\usepackage{graphicx}

\begin{document}

\title{Magneto-optical rotation of spectrally impure fields and its nonlinear dependence on optical density}
\author{G. S. Agarwal and Shubhrangshu Dasgupta}
\affiliation{Physical Research Laboratory, Navrangpura, Ahmedabad-380 009, India}
\date{\today}

\begin{abstract}
We calculate magneto-optical rotation of spectrally
impure fields in an optical thick cold atomic medium. We show that the
 spectral impurity leads to nonlinear dependence of the rotation angle on optical density. Using our calculations, we provide a quantitative analysis of the 
recent experimental results of
G. Labeyrie {\it et al.\/} [Phys. Rev. A {\bf 64}, 033402 (2001)] using cold
Rb$^{85}$ atoms..
\end{abstract}


\maketitle

\section{Introduction}
A very useful way to get important spectroscopic information is by measuring 
magneto-optical rotation (MOR) of a plane polarized light propagating through a 
medium \cite{rmp}. Clearly it is desirable to obtain as large an angle of rotation 
as possible \cite{conn,gsa}. It is known that the angle of rotation is proportional 
to the density of the medium. Thus an increase in density will help in achieving
large rotation angles. Recently, very large rotation angles in a cold sample
have been reported \cite{kaiser}. In this experiment, optical densities of the order of $10-20$ were
achieved. This experiment also reported a very interesting result, viz., the 
departure from the linear dependence of the rotation angle on the optical 
density. This departure has been ascribed to the nonmonochromatic nature of the
input laser. The findings of the experiment warrant a quantitative analysis of the
dependence of the rotation angle on the spectral profile of the input laser. We
present a first principle calculation of this dependence. Note that a nonlinear 
dependence on optical density cannot result from a simple argument based on the
standard formula for the rotation angle $\theta$:
\begin{equation}
\label{eq1}\theta=\pi kl\;\textrm{Re}(\chi_--\chi_+)\;,
\end{equation}
where, $k$ is the wave-number of the electric field, $l$ is the length of the 
medium, and $\chi_\pm$ represent the linear susceptibilities of the medium for right 
or left circularly polarized components of the input field. Since $\chi_\pm$
are proportional to the number density, the rotation angle becomes proportional to the
optical density $\alpha$ defined by 
\begin{equation}
\label{optden}\alpha=\frac{3\lambda^2}{2\pi}Nl\;,
\end{equation}
where, $\lambda=2\pi/k$ is the wavelength of the input field, $N$ is 
the number density of the atomic medium.
If one were to argue that spectrally impure laser field would replace $\chi_\pm$
by their averages over the width of the laser, then $\theta$ would continue to
be proportional to the optical density $\alpha$. A more quantitative analysis of the rotation angle is thus warranted. 

The organization of the paper is as follows. In Sec.~\ref{sec:basic}, we recapitulate the
relevant equations for studies of MOR in atomic medium.
In Sec.~\ref{sec:three}, we describe a simple three-level atomic configuration in the context of
MOR and discuss the effect of laser line-shape and magnetic field in a thick 
atomic medium. In Sec.~\ref{sec:general}, we consider experimental configuration used in \cite{kaiser}.
We show, how in an optically thick medium, one deviates from the linear dependence of rotation on optical density as a result of  
the spectral impurity of the laser field. We give a quantitative analysis of the
experimental data.

\section{\label{sec:basic}Basic equations}
Let us consider that an atomic medium of length $l$ is resonantly excited 
by a monochromatic $\hat{x}$-polarized electric field
\begin{equation}
\vec{E}(z,t)=\hat{x}{\cal E}e^{ikz-i\omega t} +\textrm{c.c.}\;,
\label{field}
\end{equation}
where, ${\cal E}$ is the field amplitude and $k=\omega/c=2\pi/\lambda$ is the 
wave number of the field, $\omega$ and $\lambda$ being the corresponding 
angular frequency and wavelength. The field is propagating in $z$-direction. 
Clearly, we can resolve the amplitude of the electric field into its two 
circular components as 
\begin{equation}
\hat{x}{\cal E}\equiv\hat{{\epsilon}}_+{\cal E}_++\hat{{\epsilon}}_-{\cal E}_-\;,
\label{circular}
\end{equation}
where, ${\cal E}_{\pm}={\cal E}/\sqrt{2}$ are the amplitude components along two
circular polarization $\hat{\epsilon}_{\pm}=(\hat{x}\pm i\hat{y})/\sqrt{2}$.

While passing through the atomic medium these two circular components 
behave differently due to the anisotropy of the medium. Let $\chi_{\pm}$ be the 
susceptibilities of the medium corresponding to the two circular components. The
electric field at 
the exit face of the medium can be written as 
\begin{eqnarray}
\vec{E}(l,t)&=&\vec{\cal E}_le^{ikz-i\omega t}+\textrm{c.c.}\;,\nonumber\\
\label{fieldout}\vec{\cal E}_l&=&\left[\hat{\epsilon}_+{\cal E}_+e^{2\pi ikl\chi_+}+\hat{\epsilon}_-{\cal E}_-e^{2\pi ikl\chi_-}\right]\;,
\end{eqnarray}
where, we have assumed that the medium is dilute so that 
$|4\pi\chi_{\pm}|\ll 1$. In MOR, the polarization 
direction of the input electric field gets rotated due to the difference in 
their dispersions (phase shifts) in a non-attenuating medium. The electric 
field however, remains linearly polarized after passing through the medium. In 
the present case, because the atom interacts with near-resonant electric 
field, the two circular components suffer attenuation (given by the imaginary 
part of $\chi_{\pm}$) while propagating through the medium. Thus the medium,
concerned here, is responsible for both dispersion and attenuation. We term 
the medium to be  both circularly birefringent and circularly dichroic. The 
output electric field becomes elliptically polarized under the action of such a medium. Thus to 
fully characterize the polarization state of the output field, 
one has to use the Stokes parameters \cite{born}. The four Stokes parameters for
an electric field are designated by $S_{\alpha}$ ($\alpha =0,1,2,3$) and can 
be defined as follows:
\begin{subequations}
\begin{eqnarray}
\label{s0}S_0&=&I_{\parallel}+I_{\perp}\;,\\
\label{s1}S_1&=&I_{\parallel}-I_{\perp}\;,\\
\label{s2}S_2&=&I_{45^o}-I_{-45^o}\;,\\
\label{s3}S_3&=&I_{\sigma_+}-I_{\sigma_-}\;,
\end{eqnarray}
\label{stokes}
\end{subequations}
where, $I_{\hat{n}}$ is the measured intensity along the polarization direction 
$\hat{n}$. Then the output polarization state
can be characterized by  the following three quantities:
\begin{subequations}
\begin{eqnarray}
\label{polar}P&=&\frac{\sqrt{S_1^2+S_2^2+S_3^2}}{S_0}\;,\\
\label{theta}\tan{2\theta}&=&\frac{S_2}{S_1}~~~~(0\leqslant\theta < \pi)\;,\\
\label{phi}\tan{2\phi}&=&\frac{S_3}{S_0P}~~~~(-\pi/4 <\phi \leqslant\pi/4)\;,
\end{eqnarray}
\label{defination}
\end{subequations}
where, $P$ is the degree of polarization, i.e., the ratio of the intensities of the 
polarized component to the unpolarized one, $\theta$ is the Faraday rotation 
angle of the input field and is measured between the major axis of the 
ellipse and the $x$-axis, $\phi$ provides the ellipticity of polarization 
through the relation $e=\tan{\phi}$.

From Eq.~(\ref{fieldout}) one can express the output intensities along 
different polarization directions in the following way:
\begin{subequations}
\begin{eqnarray}
I_{\parallel}(\omega)&=&|\hat{x}.\vec{\cal E}_l|^2=\frac{I_o}{4}|e^{2\pi ikl\chi_+}+e^{2\pi ikl \chi_-}|^2\;,\\
I_{\perp}(\omega)&=&|\hat{y}.\vec{\cal E}_l|^2=\frac{I_o}{4}|e^{2\pi ikl\chi_+}-e^{2\pi ikl\chi_-}|^2\;,\\
I_{\pm 45^o}(\omega)&=&\left|\frac{\hat{x}\pm\hat{y}}{\sqrt{2}}.\vec{\cal E}_l\right|^2\nonumber\\
&=&\frac{I_o}{8}|(1\pm i)e^{2\pi ikl\chi_+}+(1\mp i)e^{2\pi ikl\chi_-}|^2\;,\\
\label{s3a}I_{\sigma_\pm}(\omega)&=&|\hat{\epsilon}_\pm .\vec{\cal E}_l|^2=\frac{I_o}{2}\exp{[-4\pi kl.\textrm{Im}(\chi_\pm)]}\;,
\end{eqnarray}
\label{intensity}
\end{subequations}
where, $I_o=|{\cal E}|^2$ is the input intensity of laser field.

Note that all the measured quantities defined by Eq.~(\ref{intensity}) are
functions of the frequency of the exciting field. If the exciting field is
spectrally impure, then the Stokes parameters $\langle S_\alpha\rangle$ are
to be
obtained by averaging over the spectrum $S(\omega)$ of the laser field. Thus,
$I$'s in Eq.~(\ref{stokes}) are to be obtained from 
\begin{equation}
\langle I_{\hat{n}}\rangle=\frac{1}{I_o}\int_{-\infty}^{\infty}d\omega I_{\hat{n}}(\omega)S(\omega)\;.
\label{avg}
\end{equation}    
For simplicity, we can adopt, say, a Lorentzian line-shape for the input 
\begin{equation}
S(\omega)\equiv I_o\frac{\gamma_c/\pi}{\gamma_c^2+(\omega-\omega_l)^2}\;,
\label{lineshape}
\end{equation}
where, $\omega_l$ is the central frequency of the laser field and $2\gamma_c$ is the full width at half maximum. We will demonstrate how the fluctuations of the 
input field leads to the nonlinear dependence of the rotation angle on
optical density.

\begin{figure}
\scalebox{0.75}{\includegraphics{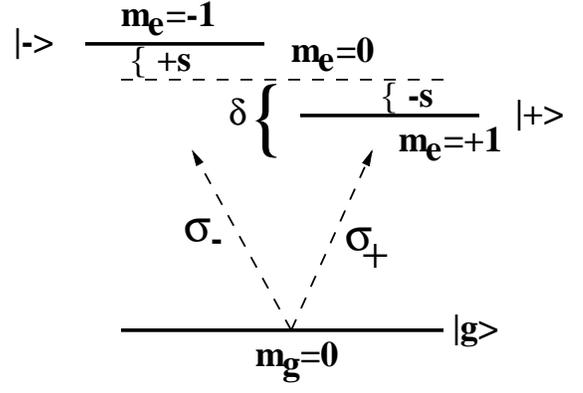}}
\caption{\label{fig1}Level diagram for a three-level configuration. The 
excited levels $|\pm\rangle$ ($m_e=\pm 1$) are Zeeman shifted from the level $m_e=0$ by an
amount $s$. The detuning $\delta$ is defined between the levels $m_e=0$ and
$m_g=0$.}
\end{figure}

\section{\label{sec:three}A simplified atomic model}
We first consider a three-level atom in $V$ configuration (see Fig.~\ref{fig1}) in order to uncover the effect of optical density on MOR. The levels $|\pm\rangle$
($J_e=1, m_e=\pm 1$) are coupled to the ground state $|g\rangle$ 
($J_g=0, m_g=0$) by two circular components $\sigma_\pm$ of $\hat{x}$-polarized
electric field [Eq. (\ref{field})]. The excited level degeneracy 
has been removed by an uniform magnetic field $\vec{B}$ applied in the 
direction of propagation of the applied electric field. The levels 
$|e_\pm\rangle$ are shifted about line-center by an amount of $\mp \mu_BB/\hbar$
($\mu_B$=Bohr magneton). The field $\vec{E}$ is detuned from the 
line center by an amount $\delta=\omega_{+g}(B=0)-\omega$, $\omega_{+g}(B=0)$ 
being the atomic transition frequency in absence of the magnetic field.

The susceptibilities of the $\sigma_\pm$ components inside the medium can be 
written as 
\begin{equation}
\chi_\pm=\frac{N|\vec{d}|^2}{\hbar\gamma}\rho_\pm;~~\rho_\pm=\frac{i\gamma}{\gamma+i(\delta\mp s)}\;,
\label{succep}
\end{equation}
where, $2\gamma=4|\vec{d}|^2\omega^3/3\hbar c^3$ is the 
spontaneous decay rate of the levels $|\pm\rangle$, $|\vec{d}|$ is the 
magnitude of the dipole moment vector for the transitions $|\pm\rangle\leftrightarrow |g\rangle$, 
$N$ is the atomic number density, and $s=\mu_BB/\hbar$ is the Zeeman 
splitting of the excited levels. Using these $\chi_\pm$, we can now write the 
field amplitude from Eq.~(\ref{fieldout}) as 
\begin{equation}
\vec{\cal E}_0=\left[\hat{\epsilon}_+{\cal E}_+e^{i\frac{\alpha}{2}\rho_+}+\hat{\epsilon}_-{\cal E}_-e^{i\frac{\alpha}{2}\rho_-}\right]\;,
\label{eout1}
\end{equation}
where, $\alpha=4\pi klN|\vec{d}|^2/\hbar\gamma=(3\lambda^2/2\pi)Nl$ is the optical density of the medium.

\begin{figure}
\scalebox{0.4}{\includegraphics{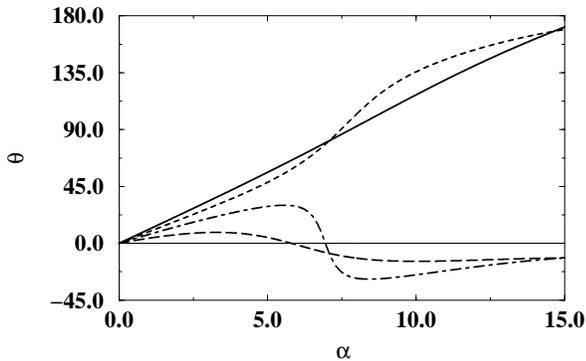}}
\caption{\label{fig2}Variation of MOR angle $\theta$ with optical density 
$\alpha$ for $s=2\gamma$ and different laser line-widths $\gamma_c=0.1\gamma$ (solid line), 
$\gamma_c=0.5\gamma$ (dashed line), $\gamma_c=\gamma$ (dot-dashed line), and $\gamma_c=2\gamma$
(long-dashed line). We have chosen $\lambda=422.67$ nm corresponding to Ca$^{40}$ 1S$_0 \leftrightarrow 1$P$_1$ transitions. Note the nonlinear dependence of $\theta$ on $\alpha$ for 
larger $\gamma_c$.}
\end{figure}

In what follows, we will assume that $\omega_l=\omega_{+g}(B=0)$. We calculate $\langle I_{\hat{n}}\rangle$'s using Eqs.~(\ref{lineshape}) and 
(\ref{avg}) numerically for different values of $\gamma_c$ and $s$. We
show the results in Fig.~\ref{fig2}. We clearly see 
that for $\gamma_c\ll s$, the rotation angle $\theta$ is 
linearly proportional to $\alpha$. But for $\gamma_c\gtrsim s$, this variation 
deviates from linearity in large $\alpha$ domain. This behavior can be explained
in terms of the off-resonant components which dominate for the large $\gamma_c$
and large $\alpha$. 

In order to understand the numerical results, we first consider the limit of
small optical densities whence
\begin{subequations}
\begin{eqnarray}
S_1&=&\frac{1}{2}[2+\alpha\textrm{Re}\{i(\rho_++\rho_-)\}]\;,\\
S_2&=&\frac{\alpha}{2}\textrm{Re}(\rho_--\rho_+)\;.
\end{eqnarray}
\label{small1}
\end{subequations}
Thus, the departure of the rotation angle from the linearity has to do with the
averages of the exponentials appearing in $I$'s [Eq.~(\ref{intensity})]. If
one were to make the approximation of replacing all $\chi$'s in Eq.~(\ref{intensity})
by their averages, i.e., 
\begin{equation}
\langle \exp{[2\pi ikl\chi_\pm]}\rangle\equiv \exp[2\pi ikl\langle\chi_\pm\rangle]\;,
\label{approx}
\end{equation}
then the Stokes parameters $\langle S_1\rangle$ and $\langle S_2\rangle$ would be
\begin{subequations}
\begin{eqnarray}
\langle S_1\rangle&=&e^{-\frac{\alpha}{2}(\langle\rho_2^+\rangle+\langle\rho_2^-\rangle)}\cos{\left[\frac{\alpha}{2}(\langle\rho_1^+\rangle-\langle\rho_1^-\rangle)\right]}\;,\\
\langle S_2\rangle&=&-e^{-\frac{\alpha}{2}(\langle\rho_2^+\rangle+\langle\rho_2^-\rangle)}\sin{\left[\frac{\alpha}{2}(\langle\rho_1^+\rangle-\langle\rho_1^-\rangle)\right]}\;,
\end{eqnarray}
\label{small3}
\end{subequations}
where, $\langle\rho_\pm\rangle=\langle\rho_1^\pm\rangle +i\langle\rho_2^\pm\rangle$ and thus
\begin{figure*}
\scalebox{0.9}{\includegraphics{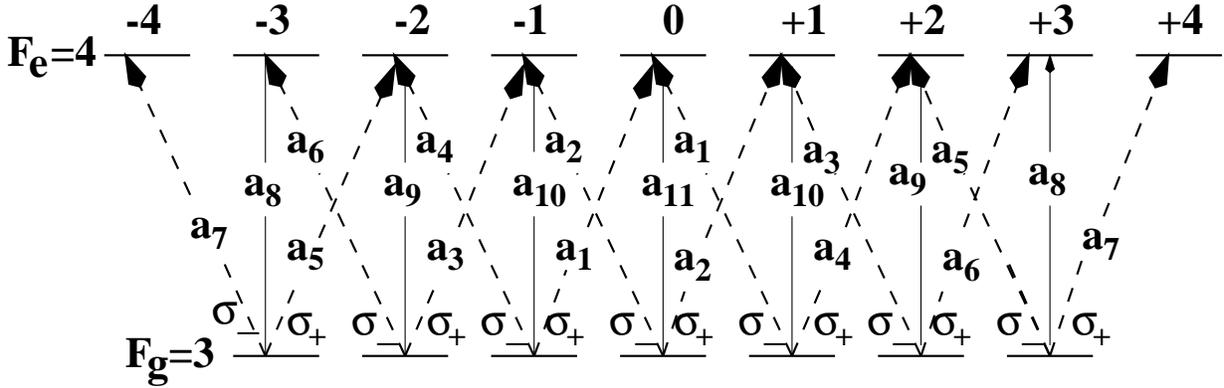}}
\caption{\label{fig16}Level diagram for the $F_e=4 \leftrightarrow F_g=3$
transition. The numbers at the top of the figure indicate the magnetic 
quantum numbers of the sublevels. The relevant Clebsch-Gordan coefficients for the corresponding 
transitions are given by $a_1=-1/\sqrt{42}, a_2=-\sqrt{5}/3\sqrt{14}, a_3=-1/2\sqrt{21}, a_4=-\sqrt{5}/2\sqrt{21}, a_5=-1/6\sqrt{7}, a_6=-1/2\sqrt{3}, a_7=-1/3, a_8=-1/6, a_9=-1/\sqrt{21}, a_{10}=-\sqrt{15}/6\sqrt{7}, a_{11}=2/3\sqrt{7}$.
The Zeeman splitting of the various sublevels are not shown.}
\end{figure*}
\begin{equation}
\langle\theta\rangle=\frac{\alpha}{4}(\langle\rho_1^-\rangle-\langle\rho_1^+\rangle)=\frac{1}{2}.\frac{\alpha\gamma s}{(\gamma+\gamma_c)^2+s^2}\;.
\label{small4}
\end{equation}
Clearly, the absorption does not contribute to the rotation angle. We have again
recovered the linear dependence of $\theta$ on $\alpha$, provided the approximation
(\ref{approx}) is valid. Thus, any departure in linearity of $\theta$ with 
respect to $\alpha$ indicates the breakdown the approximation (\ref{approx}).
The numerical results of Fig.~\ref{fig2}  clearly show the breakdown of
the
mean field description obtained by replacing $\chi$'s by their average values.
    
From the Eq.~(\ref{small4}), we readily see that in the low
$\alpha$-domain, by increasing $s$ (or $\gamma_c$) while keeping $\gamma_c$ (or $s$)
constant, the slope of $\theta$ with $\alpha$ decreases. This is clear from 
the numerical results of Fig.~\ref{fig2}.  Also, for larger
values of $\gamma_c$, variation of $\theta$ with $\alpha$ is deviated from 
linearity. 
Linear variation of $\theta$ with $\alpha$ is attributed to the monochromatic laser field. 
If the electric field is spectrally impure, then the off-resonant components  
also contribute to $\theta$, through the relations (\ref{avg}), (\ref{stokes}), and (\ref{defination}).
Thus, $\theta$ starts 
varying with $\alpha$ linearly in low $\alpha$ limit, then saturates, and finally decreases 
to zero to change the direction of rotation, for larger $\gamma_c$ (see Fig.~\ref{fig2}). But for smaller 
$\gamma_c$, the linear behavior is retained even for larger $\alpha$, as
the off-resonant components are not dominant in this parameter zone.
\begin{figure*}
\begin{tabular}{cc}
\scalebox{0.4}{\includegraphics{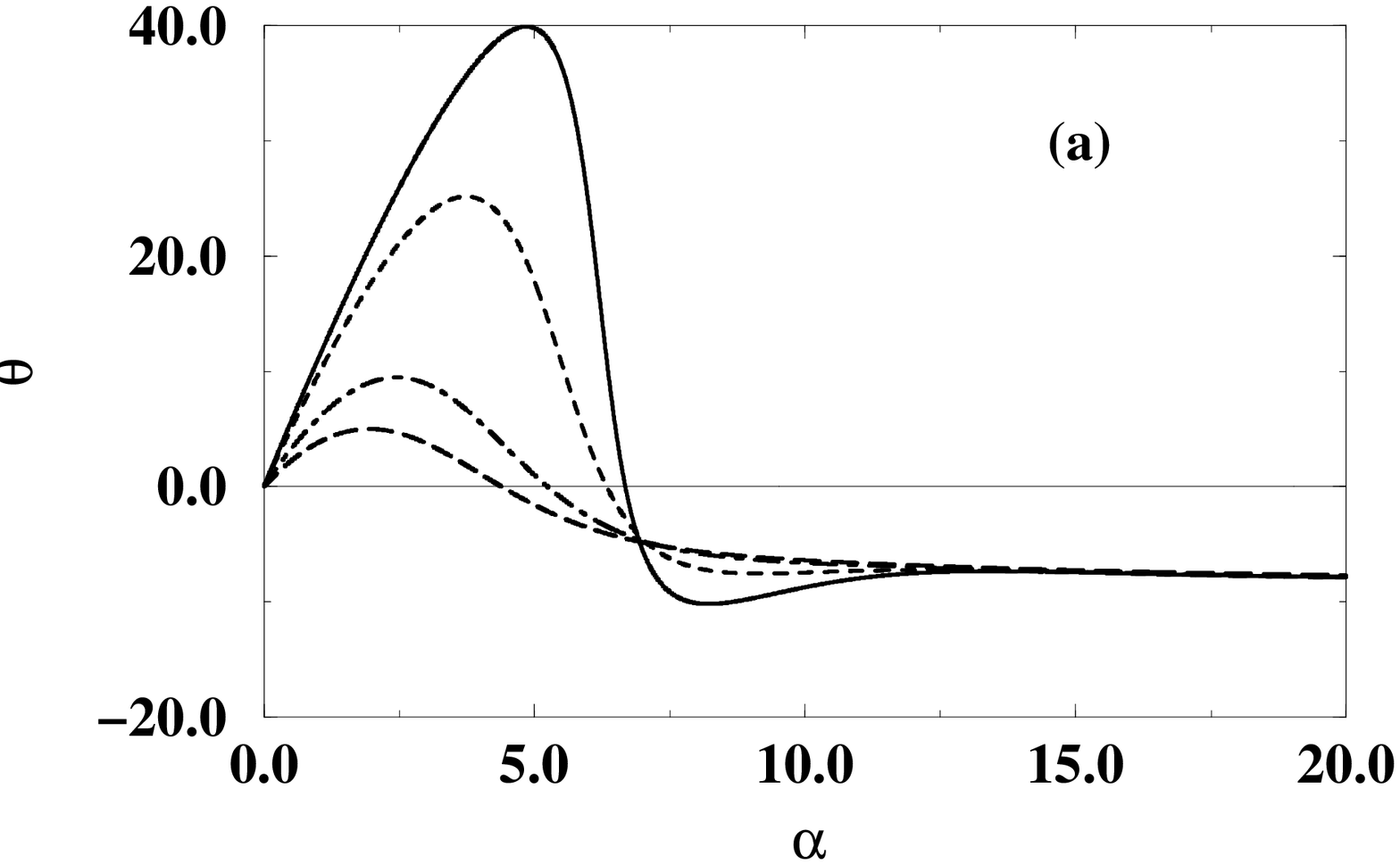}}&
\scalebox{0.4}{\includegraphics{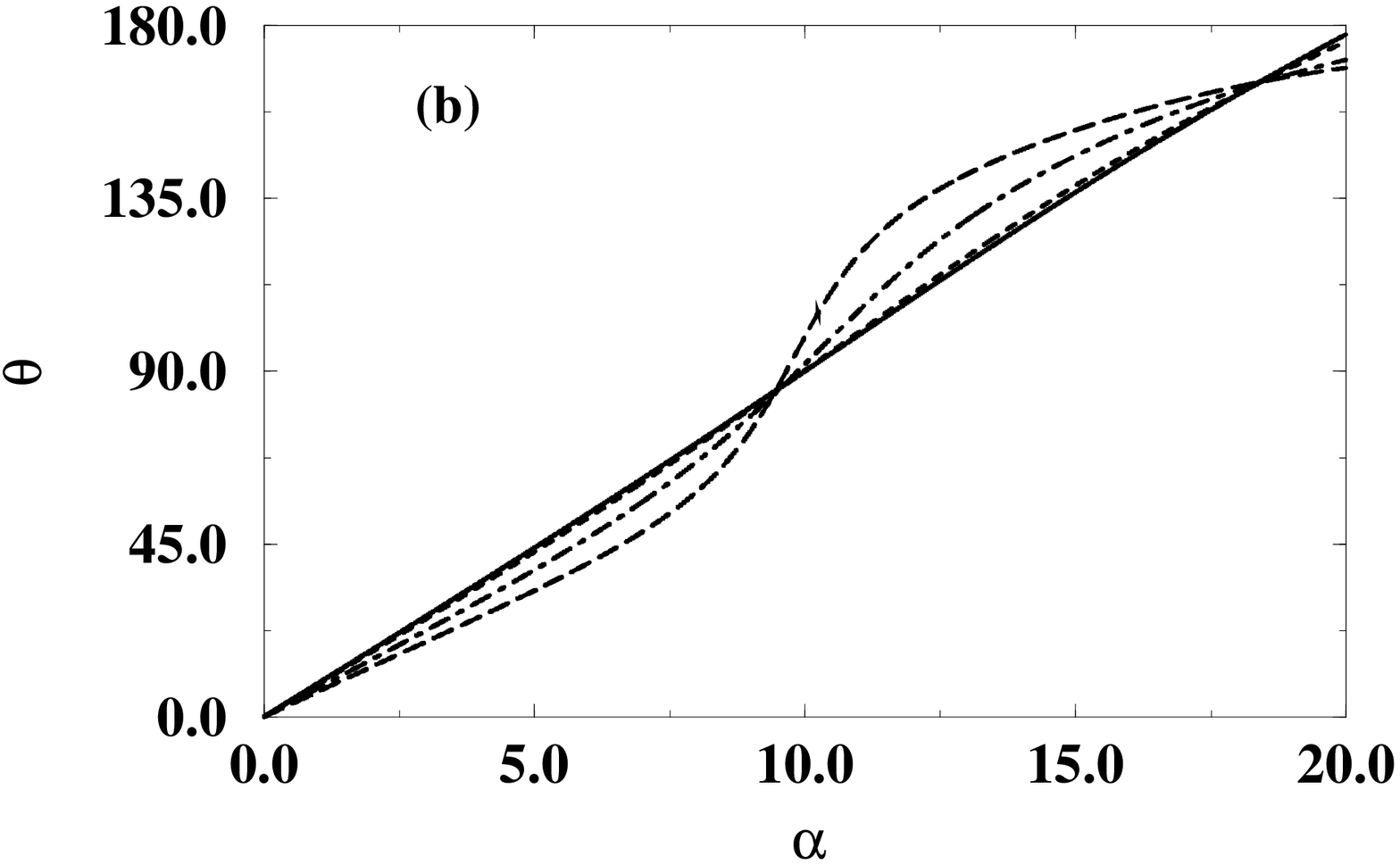}}
\end{tabular}
\caption{\label{gen}Variation of magneto-optical rotation angle $\theta$ with 
$\alpha$ for magnetic field (a) 2 Gauss ($\equiv 2\pi\times 2.8$ MHz) and (b) 8 Gauss ($\equiv 2\pi\times 11.2$ MHz) for laser line-widths
$2\gamma_c=2\pi\times 0.5$ MHz (solid line), $2\gamma_c=2\pi\times 1$ MHz (dashed line), $2\gamma_c=2\pi\times 3$ MHz (dot-dashed line), and $2\gamma_c=2\pi\times 5$ MHz (long-dashed line). The dot-dashed curves correspond to the width of the laser used in the experiment \cite{kaiser}. Note that the line-width of the D$_2$ line is $2\pi\times 5.88$ MHz.}
\end{figure*}

Next we consider the variation of the degree of polarization $P$ and the ellipticity
$e$ with $\alpha$. We have noticed that $P$ decreases from unity for increasing
$\alpha$. This means that the output field no longer remains fully polarized,
rather it becomes partially polarized. 

Again, from the Eqs.~(\ref{s3}), (\ref{s3a}), and (\ref{succep}),  it is clear 
that an integration over the entire range of detuning $\delta$ would yield 
$\langle S_3\rangle=0$, as the integrand is an odd function of $\omega$.
Thus the ellipticity $e$ becomes zero. This means that
the polarized part of the output field remains linear. 
\begin{figure*}
\begin{tabular}{cc}
\scalebox{0.4}{\includegraphics{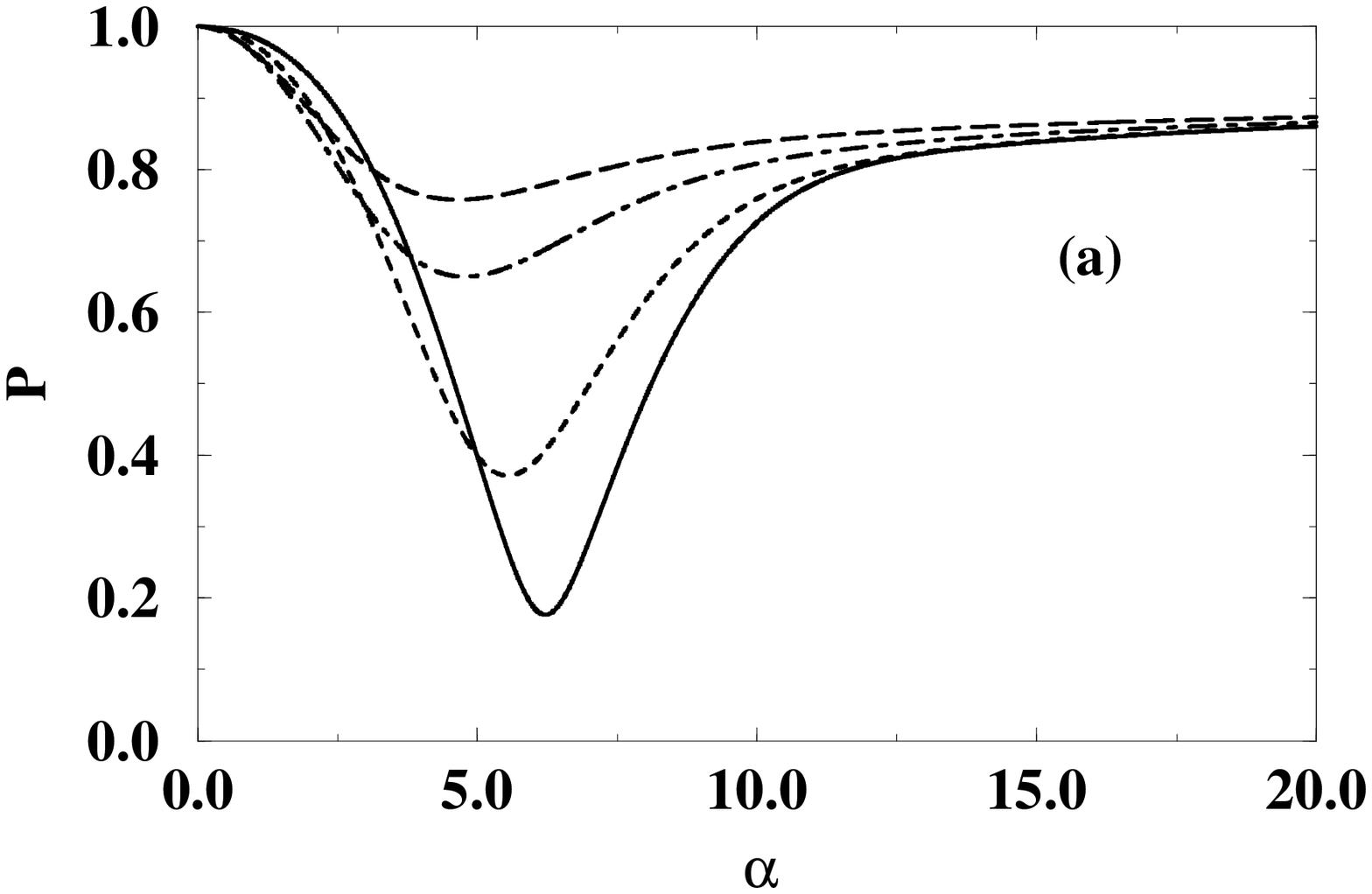}}&
\scalebox{0.4}{\includegraphics{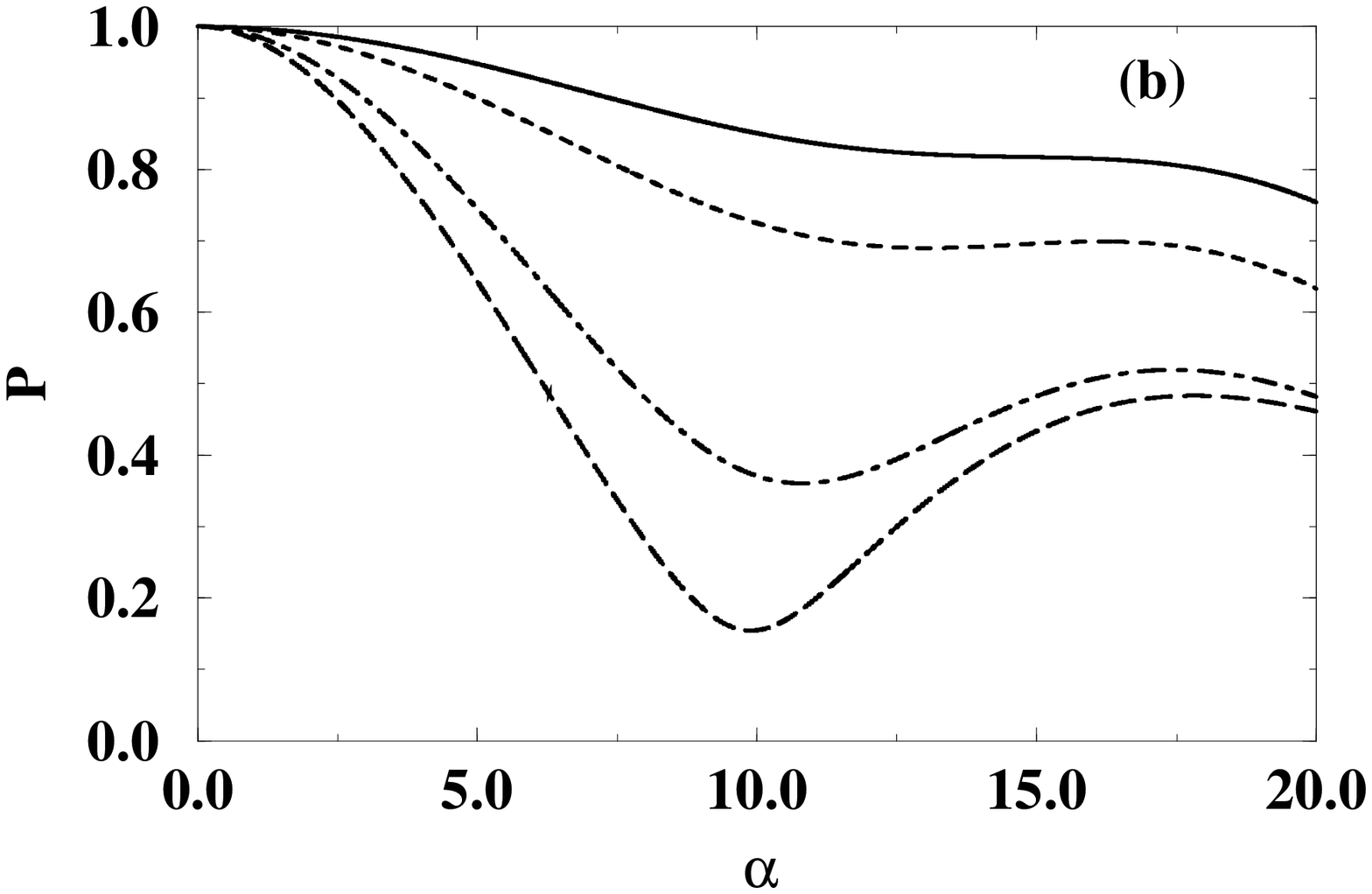}}
\end{tabular}
\caption{\label{figP}Variation of degree of polarization $P$ with $\alpha$ is shown for magnetic field
(a) 2 Gauss and (b) 8 Gauss, for laser line-widths $2\gamma_c=2\pi\times 0.5$ MHz (solid line), $2\gamma_c=2\pi\times 1$ MHz (dashed line), $2\gamma_c=2\pi\times 3$ MHz (dot-dashed line), and $2\gamma_c=2\pi\times 5$ MHz (long-dashed line). The dot-dashed curves correspond to the width of the laser used in the 
experiment \cite{kaiser}.}
\end{figure*}

From the above discussion, it is clear that the output field gets
rotated as a manifestation of cumulative effect of optical density, magnetic
field, and laser line-width. Also it becomes partially polarized with
no ellipticity.

\section{\label{sec:general}Quantitative modeling of experimental results of
Labeyrie \lowercase{{\it et al.\/}} for MOR in spectrally impure fields}

We now extend our understanding of resonant MOR as described in the previous
section to explain the experimental data of Labeyrie {\it et al.\/}. In
their experiment, a cold atomic cloud of Rb$^{85}$ is subjected to a 
static magnetic field. The laser probe beam passing through the medium in 
the direction of the magnetic field is tuned to the D$_2$ line of the atoms 
(2S$_{1/2}\leftrightarrow$ 2P$_{3/2}$; $\lambda$=780.2 nm). They have measured the
intensities of outputs with different polarizations, as function of laser
detuning and also at different values of optical density. They have found a 
nonlinear dependence of the
MOR angle $\theta$ on optical density. They found for larger magnetic field that
the linear behavior is recovered. 

To explain these observations, we consider the relevant energy levels of
Rb$^{85}$ as used in the experiment (see Fig.~\ref{fig16}). The $\hat{x}$-polarized electric field
(\ref{field}) is applied to cold Rb$^{85}$ medium near resonantly. The
medium is subjected to uniform magnetic field $\vec{B}$ applied in the $z$-direction,
i.e., along the direction of propagation of (\ref{field}).

\subsection{Calculation of $\chi_\pm$ and optical density}

The circular components $\sigma_\pm$ of the input electric field (\ref{field})
interact with the transitions $m_e\leftrightarrow m_g=m_e-1$ and 
$m_e\leftrightarrow m_g=m_e+1$, respectively. We assume that the electric field is weak enough
so that it is sufficient to use the linear response of the system to the laser
field. We neglect the ground-state
coherences. As we are considering the cold atoms, we neglect the collisional
relaxations and Doppler broadening of the sublevels. We also assume that the atomic
population is equally distributed over all the ground sublevels.

Using all these assumptions, we can write the susceptibilities $\chi_\pm$ for the 
$\sigma_\pm$ components as the sum of the susceptibilities of all the relevant 
$m_e\leftrightarrow m_g$ transitions in the following way:
\begin{equation}
\label{chis}\chi_\pm\equiv\sum_{m_e,m_g}\frac{1}{7}\frac{N|\vec{d}_{m_e,m_g}.\hat{\epsilon}_\pm|^2}{\hbar}.\frac{i}{\Gamma_{m_e,m_g}+i(\delta+ s_{m_e,m_g})}\;,
\end{equation}
where, $s_{m_e,m_g}=(g_gm_g-g_em_e)s$ is the relative amount of 
Zeeman shift of the excited sublevel $m_e$ with respect to the
Zeeman shifted ground sublevel $m_g$, $g_g=1/3$ and $g_e=1/2$ are the Land\'e-g
factors of the ground and excited levels, respectively.
The factor $1/7$ comes into the expression (\ref{chis}) as we have assumed equal population distribution in all the $(2F_g+1)=7$ ground sublevels.
The coherence 
relaxation rate $\Gamma_{m_e,m_g}$ in Eq.~(\ref{chis}) is given by 
\begin{equation}
\Gamma_{m_e,m_g}=\frac{1}{2}\sum_{k}\gamma_{k,m_e}\;,
\end{equation}
where, $\gamma_{i,j}$ is the spontaneous relaxation rate from the 
sublevel $j$ to $i$. Here we have assumed that there is 
no spontaneous relaxation from the ground sublevels. The terms $\vec{d}_{m_e,m_g}$ and $\Gamma_{m_e,m_g}$'s can be
calculated from the relevant Clebsch-Gordan coefficients (see Fig.~\ref{fig16})
\cite{sobel}. The Einstein's A coefficient for the D$_2$ line is known to be
\begin{widetext}
\begin{equation}
A=\frac{4\omega^3}{3\hbar c^3}\frac{|(J=\frac{3}{2}\parallel D\parallel J'=\frac{1}{2})|^2}{4}=\frac{4\omega^3}{3\hbar c^3}\frac{|(\frac{3}{2}, \frac{5}{2}, 4\parallel D\parallel \frac{1}{2}, \frac{5}{2}, 3)|^2}{9}\;,
\end{equation}
\end{widetext}
where, $(~\parallel D\parallel ~)$ represents the reduced matrix element of the dipole moment vector $\vec{d}_{m_e,m_g}$. The three symbols $3/2$, $5/2$, and $4$ correspond to $J$, $I$, and $F$ values respectively of the upper levels.
Thus all $\Gamma_{m_e,m_g}$'s in (\ref{chis}) are found to be equal to $(4\omega^3/3\hbar c^3)|(\frac{3}{2}, \frac{5}{2}, 4\parallel D\parallel \frac{1}{2}, \frac{5}{2}, 3)|^2/2$. 

We calculate the optical density $\alpha$ of the medium, when the input light
field is resonant with $m_e=0\leftrightarrow m_g=0$ transition ($\delta=0$) in the 
absence of any magnetic field ($B=0$). For this, we first obtain the total output 
intensity from Eq.~(\ref{s0}) 
averaged by a very narrow laser line-shape, i.e., in the limit $\gamma_c\rightarrow 0$. Using Eq.~(\ref{chis}), we thus find that the transmittivity of the 
medium becomes 
\begin{equation}
T=\frac{1}{I_o}\langle S_0\rangle_{\gamma_c\rightarrow 0}=\frac{1}{I_o}S_0|_{\delta=0}=\exp(-\alpha)\;,
\end{equation}
where, $\alpha=(3/7).(3\lambda^2/2\pi) Nl$.
It should be borne in mind that it is different from the definition in Sec.~\ref{sec:three}.

\subsection{Discussions}

Using the above expressions of $\chi_\pm$ [Eq.~(\ref{chis})] and Eq.~(\ref{avg}), 
we calculate the averaged intensities $\langle I_{\hat{n}}\rangle$ in different polarization directions. The Stokes parameters
$S_\alpha$, degree of polarization $P$, and the Faraday rotation $\theta$ are 
calculated using the relations (\ref{defination}).
In Fig.~\ref{gen}, we show how the Faraday angle $\theta$ varies with the
optical density $\alpha$ for different values of $\gamma_c$ and $B$. Clearly,
for $\gamma_c\ll s$, the rotation angle $\theta$ varies linearly with $\alpha$.
 But for larger $\gamma_c$ ($\gtrsim s$), the variation of $\theta$
with $\alpha$ deviates from linearity in large $\alpha$. This is because the 
off-resonant components contribute to the output intensity. Also, note that 
for a given value of
$\gamma_c$, if $s$ is increased, the linearity is maintained even in the large $\alpha$-domain. 
This is because for larger $s$, the off-resonant components do not contribute 
much to the output intensity. The resonant frequency component is always 
dominant in the optical density range considered. We also note that, as 
$\gamma_c$ increases, the linear slope of $\theta$ with $\alpha$ decreases in the small $\alpha$  domain.

In Fig.~\ref{figP}, we show the variation of degree of polarization $P$ with $\alpha$ for various values of the $B$ and $\gamma_c$. These results reveal that 
with increase in $\alpha$, the degree of polarization deviates from unity, 
i.e., the
output electric field not only rotates in polarization, but  also it becomes 
{\it partially\/} polarized. However, the ellipticity of the output field still
remains zero as we have argued in Sec.~\ref{sec:three}.

\section{conclusions}

In summary, we have given a quantitative analysis of magneto-optical rotation of
spectrally impure fields in optically thick cold Rb$^{85}$ atomic medium. We have shown that the dependence of rotation
on the optical density of the medium deviates from linearity 
due to the finite  laser linewidth. Using our
model, we explain the experimental results of Labeyrie {\it et al.\/}.

\end{document}